\documentclass[12pt]{article}
\usepackage{graphics}
\usepackage{amsfonts}
\usepackage{amsmath}
\usepackage{amssymb}

\def\\{\hfill\break} \let\==\equiv

\let\0=\noindent

\def\qed{\hfill\raise1pt\hbox{\vrule height5pt width5pt depth0pt}}

\def\be{\begin{equation}}
\def\ee{\end{equation}}
\def\bea{\begin{eqnarray}}\def\eea{\end{eqnarray}}

\begin{document}
\title{A Quantum Story}
\author {Stephen Boughn{\small\it\thanks{sboughn@haverford.edu}}\author{Stephen Boughn}
\\[2mm]
 \it Department of Physics, Princeton University, Princeton NJ\\
 \it Departments of Physics and Astronomy, Haverford College, Haverford PA}

\date{{\small   \LaTeX-ed \today}}

\maketitle

\begin{abstract}

After the development of a self-consistent quantum formalism nearly a century ago there began a quest for how to interpret the theoretical constructs of the formalism.  In fact, the pursuit of new interpretations of quantum mechanics persists to this day.  Most of these endeavors assume the validity of standard quantum formalism and proceed to ponder the ontic nature of wave functions, operators, and the Sch\"{o}dinger equation. The present essay takes a different approach, more epistemological than ontological. I endeavor to give a heuristic account of how empirical principles lead us to a quantum mechanical description of the world.  An outcome of this approach is the suggestion that the notion of discrete quanta leads to the wave nature and statistical behavior of matter rather than the other way around.  Finally, the hope is to offer some solace to those older of us who still worry about such things and also to provide the neophyte student of quantum mechanics with physical insight into the mathematically abstract and often baffling aspects of the theory.


\end{abstract}



\section{Introduction}\label{INTRO}

From the very beginnings of quantum mechanics a century ago, it was clear that the concepts of classical physics were insufficient for describing many phenomena.  In particular, the fact that electrons and light exhibited both the properties of particles and the properties of waves was anathema to classical physics. After the development of a self-consistent quantum formalism, there began a quest for just how to interpret the new theoretical constructs.  De Broglie and Schr\"{o}dinger favored interpreting quantum waves as depicting a continuous distributions of matter while Einstein and Born suggested that they only provide a statistical measure of where a particle of matter or radiation might be.  After 1930, the {\it Copenhagen interpretation} of Bohr and Heisenberg was generally accepted; although, Bohr and Heisenberg often emphasized different aspects of the interpretation and there has never been complete agreement as to its meaning even among its proponents\cite{St1972}.   The Copenhagen interpretation dealt with the incongruous dual wave and particle properties by embracing Bohr's {\it principle of complementarity} in which complementary features of physical systems can only be accessed by experiments designed to observe one or the other but not both of these features.   For example, one can observe either the particle behavior or wave behavior of electrons but not both at the same time.  In addition, the waves implicit in Schr\"{o}dinger's equation were interpreted as probability amplitudes for the outcomes of experiments.  Finally, in order to facilitate the communication of experimental results, the Copenhagen interpretation emphasized that the description of experiments, which invariably involve macroscopic apparatus, must be described in classical terms.

These aspects of quantum theory are familiar to all beginning students of quantum mechanics; however, many students harbor the uneasy feeling that something is missing.  How can an electron in some circumstances exhibit the properties of a particle and at other times exhibit the properties of a wave?  How is it that a primary theoretical constructs of quantum mechanics, the Schr\"{o}dinger wave functions or Hilbert state vectors, only indicate the probability of events?  Quantum mechanics itself does not seem to indicate that any event actually happens.  Why is it that experiments are only to be described classically?  Where is the quantum/classical divide between the quantum system and the classical measurement and what governs interactions across this divide?  In fact, these sorts of questions are raised not only by neophyte students of quantum mechanics but also by seasoned practitioners.  In actuality, the question of how to interpret quantum theory has never been fully answered and new points of view are still being offered.  Many of these interpretations involve novel mathematical formalisms that have proved to be useful additions to quantum theory.  In fact, new formulations of quantum mechanics and quantum field theory, including axiomatic approaches, are often accompanied by new or modified interpretations.   Such interpretive analyses are largely framed within the mathematical formalism of quantum theory and I will refrain from saying anything more about them.

The purpose of this essay is to address a different, more epistemological question, ``What is it about the physical world that leads us to a quantum theoretic model of it?"  The intention is to in no way malign the more formal investigations of quantum mechanics.  Such investigations have been extremely successful in furthering our understanding of quantum theory as well as increasing our ability to predict and make use of novel quantum phenomena.  These treatments invariably begin with the assumption that standard quantum mechanics is a fundamental law of nature and then proceed with interpreting its consequences. In this essay I take the point of view that quantum mechanics is a model, a human invention, created to help us describe and understand our world and then proceed to address the more philosophical question posed above, a question that is still pondered by some physicists and philosophers and certainly by many physics students when they first encounter quantum mechanics. Most of the latter group eventually come to some understanding, perhaps via the ubiquitous Copenhagen Interpretation, and then proceed according to the maxim ``Shut up and calculate!"\footnote{The full David Mermin quote is ``If I were forced to sum up in one sentence what the Copenhagen interpretation says to me, it would be `Shut up and calculate!'"\cite{Me1989}} One modest aim of this essay is to provide such students with a heuristic perspective on quantum mechanics that might enable them to proceed to calculations without first having to ``shut up".

\section{What's Quantized?}
Let us begin by asking where the `quantum' in quantum mechanics comes from.  What is it that's quantized?  That matter is composed of discrete quanta, atoms, was contemplated by Greek philosophers in the $5^{th}$ century B.C.\cite{Be2011} and the idea continued to be espoused through the $18^{th}$ century.  Even though it wasn't until the $19^{th}$ and early $20^{th}$ centuries that the existence of atoms was placed on a firm empirical basis, it's not difficult to imagine what led early philosophers to an atomistic model.  Perhaps the primary motivation, an argument that still resonates today, was to address the puzzle of change, i.e., the transformation of matter.  This was often expressed by the assertion that things cannot come from nothing nor can they ever return to nothing.  Rather, creation, destruction, and change are most simply explained by the rearrangement of the atomic constituents of matter. In his epic poem {\it De rerum natura} (On the Nature of Things, circa 55 BC), Lucretius\footnote{Lucretius was a disciple of the Greek atomist Epicurus and his predecessors Democritus and Leucippus\cite{Be2011}.} explained (translation by R. Melville\cite{Lu55})
\begin{quote}
...no single thing returns to nothing but at its dissolution everything returns to matter's primal particles...they must for sure consist of changeless matter. For if the primal atoms could suffer change...then no more would certainty exist of what can be and what cannot...Nor could so oft the race of men repeat the nature, manners, habits of their parents.
\end{quote}
While it took nearly 2500 years, the conjectures of the atomists were largely justified.

One might also reasonably ask, ``Are there other aspects of nature that are quantized?"  It's no coincidence that during the same period that saw the confirmation of the atomic hypothesis, there appeared evidence for the discrete nature of atomic interactions.  Perhaps the first clues were the early $19^{th}$ century observations by Wollaston and Fraunhofer of discrete absorption lines in the spectrum of the sun and the subsequent identification of emission lines in the spectra of elements in the laboratory by Kirchoff and Bunsen in 1859.   In 1888, Rydberg was able to relate the wavelengths of these discrete spectral lines to ratios of integers.  Boltzmann introduced discrete energy as early as 1868 but only as a computational device in statistical mechanics.  It was in 1900 that Planck found he must take such quantization more seriously in his derivation of the Planck black body formula\cite{Ba2009}.  A decade later Jeans, Poincar\'{e}, and Ehrenfest demonstrated that the discreteness of energy states, which source black body radiation, follows from the general morphology of the spectrum and is not the consequence of precisely fitting the observed spectral data\cite{No1993}.  In 1905 Einstein introduced the notion of quanta of light with energies that depended on frequency with precisely the same relation as introduced by Planck\footnote{It is interesting that in 1899, the year before his seminal Planck's Law paper, Planck introduced the constant that bears his name (although he gave it the symbol $b$) from Paschen's fit of spectral data to Wien's Law.  Even then he identified it as a fundamental constant of nature along side $e$, $c$, and $G$.\cite{Pl1899}}, $E=h\nu$, and then used this relation to explain qualitative observations of the photoelectric effect\footnote{Einstein's quantitative prediction was confirmed by the 1914 experiments of Robert Millikan.}.  In 1907 it was again Einstein who demonstrated that energy quantization of harmonic oscillators explained why the heat capacities of solids decrease at low temperatures.  Finally, Bohr's 1913 model of discrete energy levels of electrons in atoms explained the spectral lines of Kirchoff and Bunsen as well as resolved the conflict of Maxwell's electrodynamics with the stability of Rutherford's 1911 nuclear atomic model.

In a 1922 conversation with Heisenberg\cite{He1972}, Bohr expressed an argument for the discreteness of atomic interactions that harkened back to the ancient Greeks' arguments for atoms (and to the Lucretius quote above).  Bohr based his argument on the stability of matter, but not in the sense just mentioned.  Bohr explained, 
\begin{quote}
By `stability' I mean that the same substances always have the same properties, that the same crystals recur, the same chemical compounds, etc.  In other words, even after a host of changes due to external influences, an iron atom will always remain an iron atom, with exactly the same properties as before.  This cannot be explained by the principles of classical mechanics...according to which all effects have precisely determined causes, and according to which the present state of a phenomenon or process is fully determined by the one that immediately preceded it.
\end{quote}
In other words, in a world composed of Rutherford atoms, quantum discreteness is necessary in order to preserve the simplicity and regularity of nature.  Bohr's `stability' and Lucretius's `repeatability' clearly refer to the same aspect of nature.

It might appear from the examples given above that energy is the key dynamical quantity that must always come in discrete quanta.  However, there are problems with this demand.  For one thing, there is no fundamental constant in physics, such as the speed of light $c$, the charge on the electron $e$, Planck's constant $h$, or Newton's constant $G$, that has the units of energy.\footnote{To be sure, the Planck energy $\sqrt{\frac{\hbar c^{5}}{G}}$ is fundamental but is much too large ($2 \times 10^{9}$ Joules) to be relevant on atomic scales.} In addition, it's straightforward to demonstrate that energy is not quantized in all situations.  Suppose there were a fundamental quantum of energy $\varepsilon_{0}$ and that a free point particle already in motion acquires this much additional kinetic energy through some unspecified interaction.  When viewed by an observer that is initially co-moving with the particle, it is trivial to demonstrate that the change in its kinetic energy is significantly less than $\varepsilon_{0}$, thereby violating the hypothetical energy quantization condition.  Similar arguments show that a fundamental quantum of linear momentum is also excluded.  In fact, it's easy to think of interactions in which the energy and momentum change by arbitrarily small increments. Rutherford scattering and Compton scattering are two such examples.  Finally, we know from quantum theory that a free particle may assume any of a continuum of values of energy and momentum.

Angular momentum is another example of a dynamical quantity that might come in discrete quanta and it has the same units as $h$.  In the context of standard quantum mechanics, the quantum state of a particle can always be expressed as a linear combination of angular momentum eigenfunctions with eigenvalues that are discrete multiples of $h/2\pi$.  Furthermore, according to the standard interpretation of quantum mechanics, any measurement of the angular momentum of a system must be an eigenvalue of the angular momentum operator.  This might lead one to conclude that angular momentum somehow characterizes the quantum nature of interactions.  On the other hand, that such a specific quantity as angular momentum should occupy this primal status seems doubtful.  In addition, a procedure for how to effect a measurement corresponding to the angular momentum operator is not necessarily well-defined.  For example, as far as I know, there is no unique prescription for how to measure the angular momentum of a free particle.  One can easily conjure a sensible measurement that, depending on the chosen origin of the coordinate system, results in an arbitrary value of the angular momentum.\footnote{In fact, sensible measurements can always be found that seemingly violate the quantum hypothesis as well as the Heisenberg uncertainty principle; however, these invariably involve the inappropriate application of quantum mechanics.  Dyson \cite{Dy2002} has given several examples of such measurements.}  Finally, I remind the reader that it is not the intent of the current discussion to offer an interpretation of standard quantum mechanics but rather to understand why we are led to a quantum mechanical model of nature.  It is in this context that the semi-classical arguments excluding energy, momentum, and angular momentum have merit.

To be sure, there are many instances of the quantization of energy $E$, momentum $p$, angular momentum $L$, and even position $x$; however, the values of these quanta depend on the specifics of the system and have a wide range of values.  For example, the energy and momentum in a monochromatic beam of photons are quantized in units of $h\nu$ and $h\nu/c$ but the values of these quanta depend on the frequency of the photons and are unconstrained; they can take on any value between $0$ and $\infty$, which again argues against their primal status.  An alternate approach might be to consider the quantization of some specific combination of $E$, $p$, $L$, and $x$.  In fact, Heisenberg's indeterminacy relation\cite{He1927}, $\delta x \delta p \sim h$, points to the product of position and momentum as such a combination (and we will see that this is, indeed, the case).  It might seem inappropriate to invoke a result of quantum mechanics in the current epistemological approach; however, Heisenberg arrived at this relation, not from quantum mechanics, but rather from an empirical gedanken experiment involving a gamma ray microscope as will be discussed later.

In 1912, Nicholson proposed that the angular momentum of an electron in orbit about an atomic nucleus is quantized and the following year Ehrenfest argued that the unit of this quantum is $\hbar \equiv h/2\pi$.  In the same year, Bohr incorporated these ideas into his model of an atom, a model that provided a successful explanation of the spectrum of atomic hydrogen.  Even though $\hbar$ has the same units as angular momentum, we argued above that it is doubtful that angular momentum constitutes the fundamental quantum interaction.  Wilson, Ishiwara, Epstein, and Sommerfeld soon replaced Nicholoson/Ehrenfest/Bohr quantization with the notion that Hamilton-Jacobi action variables $J_{k}$ (for periodic systems), which also have the same units as $h$, are quantized\cite{Wh1951}.  That is, $J_{k} \equiv \oint p_{k}dx_{k} = nh$ where $p_{k}$ is the momentum conjugate to the coordinate $x_{k}$, $n$ is an integer, and the integral is taken over one cycle of periodic motion.  Recall that one rationale for excluding energy, momentum, and angular momentum as fundamental quanta is that such quantization is not necessarily independent of the frame of reference.  It is straightforward to demonstrate that the action integral is the same for all non-relativistic inertial observers, i.e., $v \ll c$, (up to an additive constant that is irrelevant for the physics of the systems)\footnote{The relativistic case is much more complicated\cite{Ki1993}; however, even then there are individual cases in which action angle variables are Lorentz invariant\cite{Tr1999}.  Fock\cite{Fo1966} demonstrated that the geodesic equation of general relativity can be expressed in terms of the Hamilton-Jacobi equation and action function S.}.   What other properties of action variables might single them out as amenable to quantization?

In 1846, the French astronomer Delaunay invented a method of solving a certain class of separable\footnote{Separable refers to problems for which Hamilton's principle function $S$ can be written as a sum of functions, each depending on a single generalized coordinate, i.e., $S= \sum S_k(q_k)$.}
 dynamical problems for which the action variables $J_{k}$ are constants of the motion\cite{La1970}.  In 1916, Ehrenfest added to the primacy of action variables by pointing out that all of these quantities that had been previously quantized were adiabatic invariants\cite{Wh1951}.  That is, they remain constant during adiabatic changes of the parameters that define a system. This implies, for example, that if one slowly changes the length of a pendulum, it will remain in the same quantum state.  This requirement is desirable; otherwise, such an adiabatic perturbation is apt to leave the system in a state that is not consistent with the quantization condition for the changed system. On the other hand, the quantization of action depends on the coordinates employed, which again seems unacceptable for a fundamental principle.  

It was Einstein who in 1917 gave a new interpretation of the quantization conditions by demonstrating that they followed from the requirement that Hamilton's principle function $S$ is multivalued such that the change in $S$ around any closed curve in configuration space is an integer times Planck's constant, i.e., $\oint p_{k}dq_{k} = \oint \frac{\partial S}{\partial q_{k}}dq_{k} = \oint dS = nh$\cite{St2005}.  This geometric expression no longer depends on the choice of coordinates.
Consider the case of a one-dimensional harmonic oscillator such that $\oint pdx = \oint p\dot{x}dt =\oint 2Kdt = 2\langle K \rangle \tau = E\tau = nh$ where $\langle K \rangle$ is the average kinetic energy, $E$ is the total energy, and $\tau$ is the period of the oscillator. Then $\Delta E=h/\tau=h\nu$, Planck's ansatz for the quantization of the energy exchanged between blackbody radiation and hypothetical harmonic oscillators of frequency $\nu=1/\tau$ embedded in the walls of a blackbody cavity.  For the case of an electromagnetic wave of wavelength $\lambda$ and frequency $\nu=c/\lambda$, the quantization condition becomes $\oint pdx=p\lambda=pc/\nu=nh$.   The energy and momentum of an electromagnetic wave are related by $E=pc$, therefore, $E=nh\nu$, Einstein's relation for photons, the quanta of electromagnetic radiation.  The same quantization rule applies to the angular momentum $p_{\phi}$ of an electron in a hydrogen atom, $\oint p_{\phi} d\phi=2\pi p_{\phi} =nh \Rightarrow p_{\phi}=n\hbar$, the Nicholoson/Ehrenfest/Bohr quantization condition.

{\it Disclaimer:}  The quantization conditions of Hamilton-Jacobi action were part of the foundations of the old quantum theory from the years 1900-1925 and have long since been superseded by the formalism of modern quantum mechanics and quantum field theory.\footnote{Although, it is possible to derive a similar set of quantization conditions from today's quantum
mechanics, i.e., $\oint \sum p_kdq_k = (n + m/4)h$, where $n$ is an arbitrary integer and $m$ is an integer related to the
caustic structure of $S$.\cite{St2005,Ke1958}} In addition, the quantization procedure fails for non-integrable (chaotic) systems.\footnote{Einstein already indicated such a problem in 1917 but it wasn't until years later that its significance to quantum mechanics, in the context of quantum chaos, was realized. Gutzwiller later demonstrated that despite the absence of a canonical quantization scheme for such cases ``strong classical-quantum correspondences exist even for chaotic systems."\cite{St2005}}  Finally, the above arguments are semi-classical and, as such, it's difficult to imagine how they can provide a firm foundation for modern quantum theory.  However, the reader should be reminded that the purpose of this essay is not an axiomatic derivation of quantum mechanics from fundamental principles but rather to acquire insight into the quantum world and thus address the question, ``What is it about the physical world that led us to a quantum theoretic model of it?"  I now continue with this task.

\section{Quantization and Waves}\label{QandW}

In 1923 Duane\cite{Du1923}, Breit\cite{Br1923}, and Compton\cite{Co1923} applied the quantization condition to the interaction of x-ray photons with an infinite, periodic crystal lattice and were able to obtain Bragg's law of reflection without directly invoking the wave nature of x-rays.  A somewhat simpler case is that of photons incident on an infinite diffraction grating.  Figure 1 is a replica of the schematic diagram in Breit's 1923 paper where $h\nu/c =p_{\gamma}$
is the momentum of the incident photon, $G$ is the diffraction grating, $D_{0},D_{\pm1},D_{\pm2},...$ are the positions of the slits of the grating, $\theta$ is the scattering angle, and $P$ is the transverse momentum of the emergent photon. Now assume that the momentum transferred from the
\begin{figure}[htb]
\vbox{\hfil\scalebox{1.0}
{\includegraphics{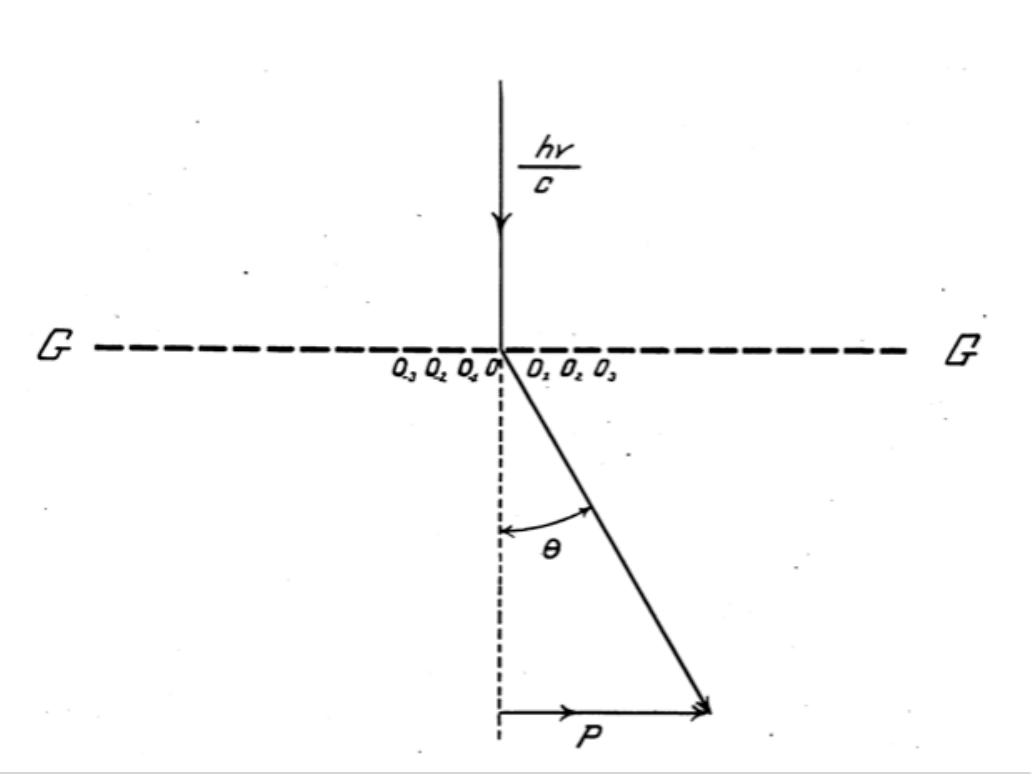}}\hfil}
{\caption{\footnotesize{Photon scattered from an infinite diffraction grating from Breit\cite{Br1923}}}}
\end{figure}
radiation to the grating is governed by $\oint pdx=nh$ where $p$ is in a direction parallel to its surface and the integral is taken over the transverse distance necessary to bring the system back to its original condition, i.e., the line spacing $d=D_{k+1}-D_{k}$.  In this case, the average momentum transferred to the grating is $\langle p \rangle=\oint pdx/\oint dx = nh/d$ and by conservation of momentum this must also be the magnitude of the transverse momentum transferred to an incident photon, i.e., $P=\langle p \rangle$.  If photons are incident perpendicular to the plane of the grating, then the allowed angles at which they are transmitted through the grating are given by $\sin \theta_n=\langle p \rangle/p_{\gamma}$.\footnote{Because the mass of the grating is very large, the momentum and energy of the scattered photon, $p_{\gamma}$ and $p_{\gamma}c$, do not change, a result that follows from the Compton effect discovered only a few months earlier.}   Thus, $\sin \theta_n=nh/p_{\gamma}d$, which is the relation for diffraction (interference) of a wave with wavelength $\lambda=h/p_{\gamma}$.  Again, no specific reference to the wave nature of the photons is necessary.  Breit\cite{Br1923} and Epstein \& Ehrenfest\cite{Ep1924} extended these results to finite width, single and multiple slit interference patterns.  Thus, the quantization condition $\oint pdx=nh$ leads directly to the interference properties of photons without directly invoking their wave nature.  It is curious that none of these authors extended their analyses to the case of electrons scattered from crystals, a process that should obey the same quantization condition.  If they had, they might have predicted that $\lambda=h/p$ and the wave nature of electrons prior to de Broglie's 1924 thesis and Davisson \& Germer's and Thomson's 1927 electron diffraction experiments. The analyses of Duane {\it et al.} provide seminal illustrations of a direct path from the quantization of action to the wave behavior of particles and photons.  As such, they lend credence to the notion that there is a primal relation between the quantization of dynamical properties and the dual wave-particle behavior of quantum systems.

\section{Physics and Probability}\label{prob}

Another major conundrum of quantum mechanics is the fundamental role of probability in the theory.\footnote{It was Einstein who first suggested that the intensity of electromagnetic waves was a measure of the probability of the location of photons. Born extended this notion to particles with a similar interpretation of the wave functions of Schr\"{o}dinger's equation.\cite{St2013}} The probabilities are taken to apply to the outcomes of possible observations of a system even though some of the observations are mutually exclusive (Bohr's {\it principle of complementarity}).  This seems to fly in the face of our classical notion that physical systems should be completely describable in isolation, prior to and independent of any observation.  How is it that the specification of mere probabilities can possibly constitute a fundamental description of a physical system and if so, how can such a description possibly provide a complete description of reality?\footnote{In fact, Einstein, Podolsky, and Rosen\cite{Ei1935} maintained ``that the description of reality as given by a [quantum] wave function is not complete."}

In 1927 Heisenberg proposed the indeterminacy relation, $\delta x\delta p \sim h$, that bears his name.  It was his contention that ``this indeterminacy is the real basis for the occurrence of statistical relations in quantum mechanics."\cite{He1927}   He arrived at the concept by considering a gedanken experiment in the form of a gamma ray microscope.  Heisenberg reasoned that with such a microscope one could only determine an electron's position to within on the order of one gamma ray wavelength, $\delta x \sim \lambda$.  But in doing so, one would impart to the electron an unknown momentum on the order of the momentum of the incident gamma ray, $\delta p \sim  E_{\gamma}/c=h\nu/c=h/\lambda$, and hence, $\delta x\delta p \sim h$.\footnote{Heisenberg also argued that similar indeterminacy relations occurred for all conjugate pairs of observable quantities.} To the extent that the wave behavior of gamma rays follows from quantization, as demonstrated by Duane {\it et al.}, the Heisenberg indeterminacy relation is a direct consequence of the quantum of action.  Heisenberg also demonstrated that this relation can be determined directly from the formalism of quantum mechanics; however, our point here is that it is already evident from the quantization of action.

Heisenberg's uncertainty principle is one of the pillars of modern physics and his gamma ray microscope provides a particularly intuitive interpretation of the principle.  However, there are other insightful gendanken experiments that are more directly tied to quantization.  For example, suppose a particle is confined to be within a one-dimensional box (potential well) of width $\ell$ but is otherwise free, i.e., has constant momentum $p$ along the one dimension but in either direction.  The motion of the particle will clearly be periodic with a spatial period $2\ell$ and the quantization condition is $\oint pdx=2p\ell=nh$.  If the particle is in its ground state, $n=1$ and $2p\ell=h$.  At any instant, the uncertainty in the particle's position is clearly $\delta x\sim \ell$.  The magnitude of the particle's momentum is known but it could be moving in either direction so the uncertainty in its momentum is $\delta p \sim h/\ell$.  Combining these two relations, we arrive at Heisenberg's indeterminacy relation, $\delta x\delta p \sim h$.  Of course, this particle is confined; however, if the box is opened, the particle is free to move in either direction.  Immediately after the box is opened, the uncertainties in the position and momentum of the now free particle again satisfy the Heisenberg relation, $\delta x\delta p \sim h$.

The argument that Heisenberg gave to support his contention that the uncertainty relations are the basis for the statistical relations in quantum mechanics is  as follows,\cite{He1927}
\begin{quote}
We have not assumed that quantum theory--in opposition to classical theory--is an essentially statistical theory in the sense that only statisitical conclusions can be drawn from precise statistical data....Rather, in all cases in which relations exist in classical theory between quantities which are really exactly measurable, the corresponding exact relations also hold in quantum theory (laws of conservation of momentum and energy).  But what is wrong in the sharp formulation of the law of causality, ``When we know the present precisely, we can predict the future," is not the conclusion but the assumption.  Even in principle we cannot know the present in all detail.  For that reason everything observed is a slection from a plenitude of possibilities and a limitation on what is possible in the future.
\end{quote} 
  
Another reason to concede to a statistical view of nature is the realization that this notion is not particularly foreign to classical physics.  Certainly, statistical mechanics is one of the triumphal successes of classical physics.  On the experimental side, careful consideration of uncertainties is always essential when comparing observations with theoretical predictions, either quantum or classical.  In the classical case these uncertainties are usually viewed as experimental ``noise" and left to the experimentalist to elucidate.  However, this doesn't necessarily have to be the case. The Hamilton-Jacobi formalism provides an approach in which such uncertainties can be included in the fundamental equations of classical mechanics\cite{Ha2005,HR2016}; although, it is usually far more convenient to deal with them in the analysis of a measurement rather than as fundamental facet of the theory.\footnote{In fact, some experimental uncertainties are routinely included in quantum mechanical calculations expressed as the weightings in {\it mixed states}.} An interesting aside is that by combining the statistical Hamilton-Jacobi formalism of classical mechanics with the Heisenberg uncertainty relations, one can generate a plausible route to Schr\"{o}dinger's equation and the concomitant wave nature of particles\cite{Ha2002,Bo2017}.   
One can even construe statistical relations in classical physics in terms of classical indeterminacy relations $\delta x > 0$ and $\delta p > 0$\cite{Vo2011}. In a very real sense, violations of these relations, namely $\delta x=0$ or $\delta p=0$, are just as inaccessible as a violation of the quantum mechanical uncertainty principle, $\delta x\delta p < \hbar/2$, an assertion to which any experimentalist will attest.\footnote{Note that here I've replaced Heiseberg's $\delta x \delta p \sim h$ with the usual $\delta x \delta p \geq \hbar/2$, which is derived from the corresponding quantum mechanical commutation relation.} These arguments are certainly not intended to demonstrate that quantum mechanics and classical mechanics are compatible.  Clearly, they are not.  They are offered simply to emphasize that probability and statistics are fundamental to physics, both classical and quantum.  Rather, the crucial difference between the two is the quantization of action that is primal in quantum physics but absent in classical physics.

\section{The Quantum/Classical Divide}\label{QCD}

The dual wave-particle nature of matter and radiation and the probabilistic nature of the theory are not the only elements that exasperate beginning students of quantum mechanics.  Another point of discomfort is the quantum/classical divide that the Copenhagen interpretation places between a quantum system and a classical measuring apparatus.  Where is the divide and what physical interactions occur at the divide? This dilemma is predicated upon the supposition that experiments must be, or inevitably are, described by classical physics.  Upon closer inspection, the assertion that classical physics adequately describes experiments is far from obvious.  Bohr expressed the situation as follows\cite{Bo1958}:
\begin{quote}
The decisive point is to recognize that the description of the experimental arrangement and the recordings of observations must be given in plain language, suitably refined by the usual terminology. This is a simple logical demand, since by the word `experiment' we can only mean a procedure regarding which we are able to communicate to others what we have done and what we have learnt.
\end{quote}
Stapp\cite{St1972} chose to emphasize this pragmatic view of classicality by using the word {\it specifications}, i.e., 
\begin{quote}
Specifications are what architects and builders, and mechanics and machinists, use to communicate to one another conditions on the concrete social realities or actualities that bind their lives together. It is hard to think of a theoretical concept that could have a more objective meaning. Specifications are described in technical jargon that is an extension of everyday language. This language may incorporate concepts from classical physics. But this fact in no way implies that these concepts are valid beyond the realm in which they are used by technicians.
\end{quote}
The point is that descriptions of experiments are invariably given in terms of operational prescriptions or specifications that can be communicated to technicians, engineers, and the physics community at large.  The formalism of quantum mechanics has absolutely nothing to say about experiments.

There have been many proposed theoretical resolutions to the problem of the quantum/classical divide but none of them seem adequate (e.g., \cite{Bo2013}).  One obvious approach is simply to treat the measuring apparatus as a quantum mechanical system.  While perhaps impractical, no one doubts that quantum mechanics applies to the bulk properties of matter and so this path might, in principle, seem reasonable.  However to the extent that it can be accomplished, the apparatus becomes part of the (probabilistic) quantum mechanical system for which yet another measuring apparatus is required to observe the combined system.  Heisenberg expressed this in the extreme case, ``One may treat the whole world as one mechanical system, but then only a mathematical problem remains while access to observation is closed off."\cite{Sch2008}

Ultimately, the dilemma of the quantum/classical divide or rather system/experiment divide is a faux problem.  Precisely the same situation occurs in classical physics but apparently has not been considered problematic.  Are the operational prescriptions of experiments part and parcel of classical theory? Are they couched in terms of point particles, rigid solid bodies, Newton's laws or Hamilton-Jacobi theory?  Of course not. They are part of Bohr's ``procedure regarding which we are able to communicate to others what we have done and what we have learnt."  Therefore, it seems that the problem of the relation of theory and measurement didn't arise with quantum mechanics but exists in classical mechanics as well.  At a 1962 conference on the foundations of quantum mechanics, Wendell Furry explained\cite{Fu1962}
\begin{quote}
So that in quantum theory we have something not really worse than we had in classical theory. In both theories you don't say what you do when you make a measurement, what the process is. But in quantum theory we have our attention focused on this situation. And we do become uncomfortable about it, because we have to talk about the effects of the measurement on the systems....I am asking for something that the formalism doesn't contain, finally when you describe a measurement. Now, classical theory doesn't contain any description of measurement. It doesn't contain anywhere near as much theory of measurement as we have here [in quantum mechanics]. There is a gap in the quantum mechanical theory of measurement. In classical theory there is practically no theory of measurement at all, as far as I know. 
\end{quote}
At that same conference Eugene Wigner put it like this \cite{Wi1962}
\begin{quote}
Now, how does the experimentalist know that this apparatus will measure for him the position? ``Oh", you say, ``he observed that apparatus. He looked at it." Well that means that he carried out a measurement on it. How did he know that the apparatus with which he carried out that measurement will tell him the properties of the apparatus? Fundamentally, this is again a chain which has no beginning. And at the end we have to say, ``We learned that as children how to judge what is around us."  And there is no way to do this scientifically. The fact that in quantum mechanics we try to analyze the measurement process only brought this home to us that much sharply.  
\end{quote}
Because physicists have long since become comfortable with the relation between theory and measurement in classical physics, perhaps the quantum case shouldn't be viewed as particularly problematic.  

\section{Back to Quanta} \label{BTQ}

I began this essay with the question ``What is it about the physical world around us that leads us to a quantum theoretic model of it?" and have tried to answer it by discussing the quantal character of the physical world along with the inevitability of the statistical nature of both quantum and classical physics.  In addition, when compared with its classical counterpart, the relation of theory and measurement in quantum mechanics doesn't seem all that unusual.  I hope these musings will provide some comfort to beginning students of quantum mechanics by providing at least a heuristic answer that bears on the epistemological origin of the dual wave-particle nature, the probabilistic interpretation of the quantum formalism, and the somewhat elusive connection of theoretical formalism and measurements.  Perhaps they will be afforded some solace as their credulity is strained by references to {\it wave-particle duality}, the {\it collapse of the wave function}, and the {\it spooky action at a distance} of entangled quantum systems.  I personally suspect that the quagmire to which we are led by these issues is spawned by conflating the physical world with the mathematical formalism that is intended only to model it, but this is a topic for another conversation.

The purpose of this essay is neither to demystify quantum mechanics nor to stifle conversation about its interpretation.  To be sure, the number of extraordinary quantum phenomena seems to be nearly without limit.  Quantum spin, anti-matter, field theory, gauge symmetry, the standard model of elementary particles, etc., are all subsequent developments in quantum theory that have very little connection to classical physics and about which the above discussion has little to say.  Certainly wave-particle duality is a mysterious fact of nature.  Whether one considers it to be a fundamental principle, as did Bohr, or sees it as intimately related to the quantal character of the world is, perhaps, a matter of taste. I have sought to couch the discussion not in the mathematical formalism of quantum theory, but in terms of a simple physical principle: {\it Matter, radiation, and their interactions occur only in discreet quanta.}  Rather than quashing discussion of the meaning of quantum mechanics, perhaps this essay will stimulate new discussions.\vspace{7 mm}

\textbf{Acknowledgements:}
I would like to thank Marcel Reginatto for many helpful conversations as well as for commenting on several early versions of this paper.  Also, thanks to Serena Connolly for introducing me to Lucretius's wonderful poem.

\end{document}